\documentclass[11pt]{article}

\usepackage[a4paper,margin=1in]{geometry}
\usepackage[T1]{fontenc}
\usepackage[utf8]{inputenc}
\usepackage{lmodern}
\usepackage{authblk}
\usepackage[hidelinks]{hyperref}

\title{Fundamental Market Design as a Layer of AI-Agent Alignment\thanks{Accepted at the \textbf{EC'26 Workshop on Incentive-Based AI Alignment}, co-located with the \textbf{27th ACM Conference on Economics and Computation}, Rome, Italy, 2026.\\[1mm]
\emph{Authors' email addresses:} \texttt{omar.inverso@gssi.it}, \texttt{emilio.tuosto@gssi.it}, \texttt{dragisa.zunic@gssi.it}.
}}

\author[1]{Omar Inverso}
\author[1]{Emilio Tuosto}
\author[1,2]{Dragi\v{s}a \v{Z}uni\'c}

\affil[1]{Gran Sasso Science Institute, Italy}
\affil[2]{Institute for Artificial Intelligence of Serbia}

\date{}

\begin{document}

\maketitle

\begin{abstract}
This paper argues that AI-agent alignment in markets should not be understood only as a property of agents, but also as a property of the interaction infrastructure in which agents act. In financial markets, this infrastructure is the market core: the rule system that determines how orders enter, interact, match, persist, and stabilize. If this fundamental interaction layer allows or rewards undesired behaviour, then higher-level alignment of agents may be insufficient.

We propose to view fundamental market design as a layer of AI-agent alignment. Alongside the important work of computational economics in modelling agents, strategies, and learning, we focus on a complementary but more fundamental layer: the formal modelling of the market core itself. Market design, especially at the level of the core mechanism, can benefit from a rigour characteristic of theoretical computer science. This gives a transparent-box model of the market, whose core properties can be formally specified and reasoned about. It also lets us treat the trading venue not as a static order book, but as a computational process combining resident orders with incoming order flow, and ask which computational model, perhaps yet unknown, naturally lies at its core.

This perspective is especially relevant for markets populated by adaptive or AI agents. Such agents may learn what the mechanism rewards, including speed, delay, liquidity provision, or manipulation. These behaviours are not only properties of individual agents, but may emerge from the agent--mechanism system. We therefore argue that transparent formal models of market cores can support incentive-oriented analysis and the design of mechanisms in which desirable behaviours are structurally favoured and undesirable behaviours are harder to sustain.
\end{abstract}

\section{Market mechanisms as alignment infrastructure}

This paper proposes a formal symbolic view of core market mechanisms as a foundation for studying AI agents as market participants. Financial markets involve heterogeneous actors -- regular traders, liquidity providers, high-frequency firms, institutional investors, and algorithmic or AI-based systems -- but all of them interact through the same rule-governed infrastructure: the trading venue itself. Like a game of chess, a market mechanism defines admissible actions, interaction rules, priority, and outcomes. For this reason, the behaviour of market participants cannot be understood only by modelling their strategies or objectives; it also depends on the formal structure of the mechanism through which they interact.

The position of this paper is that alignment is not only a property of agents, but also a property of the interaction infrastructure in which agents act. In financial markets, this infrastructure is the market core: the rule system that determines which orders may interact, which interactions have priority, how matching is performed, what persists, and when the market has reached a stable state. This perspective is especially relevant for incentive-based AI alignment, because the market core can make some behaviours natural, profitable, and stable, while making others fragile, costly, or impossible. In this sense, market design is an integral part of the alignment environment for AI agents: the core that guides their interaction can make desired behaviours easier to obtain, and undesired behaviours harder to sustain.

This suggests a layered view of alignment in markets. At higher levels, one may study agent objectives, learning rules, strategies, equilibria, welfare, and incentives. Beneath these levels, however, there is a more fundamental interaction layer: the market core, which defines order entry, matching, priority, timing, persistence, and stable states. Incentive misalignment in markets may therefore arise not only from the objectives of agents, but from the rule system through which agents interact. If this fundamental interaction layer allows or rewards latency races, strategic delay, or manipulation of timing and priority, then higher-level alignment of agents may be insufficient.

Electronic financial markets are a useful domain for this argument because they are already complex, open, and agent-based systems. Similar issues arise in other collective systems, where formal methods have been used to model and analyse how global behaviour emerges from repeated local interactions among many agents (De Nicola et al.~\cite{denicola2023}). Markets have the same character. Agents submit, cancel, and modify orders; they react to public information, visible liquidity, prices, and the rules of the trading mechanism; and by acting, they change the environment that other agents observe. The market is therefore not only a place where agents act, but also a feedback system in which agent behaviour, market state, and mechanism rules influence one another. This view is also consistent with the classic zero-intelligence trader result of Gode and Sunder~\cite{gode1993}, which showed that important market-level outcomes can arise from the structure of the market mechanism itself, even when individual agents are extremely~simple.

\section{Formalising the market core}

By ``transparency'', we do not mean only the public availability of market data, but also that the interaction rules of the market core should be explicit and formally defined. This shifts attention to agents' behaviour \emph{in the context} of well-defined (in a formal sense) rules of interaction.
The point is not to replace economic analysis, but to complement it at the level of formal design. Economics identifies incentives, welfare goals, and possible failure modes. AI studies adaptive and strategic agents. Formal methods provide tools for specifying the interaction mechanism, reasoning about its possible executions, and, where possible, building desirable properties into the core of the system rather than only observing behaviour after deployment.

This is where the computer-science contribution can be essential. Budish et al.~\cite{budish2015} have already shown that undesired strategic behaviour can arise from the structural realm itself. Formal methods, concurrency theory, operational semantics, and resource-sensitive models provide ways to specify how orders enter, interact, persist, and stabilize. These concepts make it possible to ask not only what outcomes a mechanism produces, but also which behaviours are enabled or ruled out by construction. This is a difficult design problem because financial markets are open, heterogeneous, concurrent systems with strategic agents and emergent behaviour. Precisely for that reason, market design at the fundamental level should be treated with a degree of rigour characteristic of theoretical computer science.

To make this perspective more concrete, we use Reaction Systems as a formal framework introduced by Ehrenfeucht and Rozenberg~\cite{ehrenfeucht2007}, building on our ongoing work (Brodo et al.~\cite{brodo2026}). In this setting, buy and sell orders are treated as resources; compatible orders react to produce trades; more competitive orders inhibit less competitive matches; and unmatched orders remain available for later steps. Thus, priority, competition, timing, persistence, and stability become part of the formal description of the market mechanism, rather than hidden details of implementation. This also makes key properties -- such as priority preservation, persistence of unmatched orders, and absence of further executable trades -- expressible at the level of the mechanism itself. The approach continues a line of work in which trading systems are treated as formal computational objects, rather than only as economic or statistical processes (Cervesato et al.~\cite{cervesato2019}).

A related industry-facing example is Imandra's work on trading infrastructure. In a white paper submitted to the SEC, Ignatovich and Passmore~\cite{ignatovich2015sec} argue that the complexity of modern financial algorithms calls for tools based on mathematical logic and automated reasoning. Their related work presents Imandra as a framework for reasoning about exchange and dark-pool matching logic, as well as other forms of financial infrastructure~\cite{passmore2017}.

A second precedent comes from financial contracts. Work on compositional contract languages showed that programming-language ideas and formal semantics can describe complex financial instruments precisely (Peyton Jones et al.~\cite{peytonjones2000}). These ideas influenced systems such as LexiFi, which reports that its technology, including the contract-description language, was integrated into Bloomberg's Derivatives Library~\cite{lexifiBloomberg}. 

It is useful to distinguish several classes of properties. Some are core mechanism properties: they follow directly from the design of the market model. For instance, after processing the relevant orders, the market should not be left in a crossed state, and trades should respect the priority rules of the mechanism. These properties are closest to formal specification and verification. Other properties are strategic: they concern how agents exploit or react to the rules. Latency arbitrage, sniping, front-running, strategic delay, and manipulation of priority are examples. These are not simple invariants of the market core, but they are induced by the rules of the game. A further class consists of market-quality properties, such as liquidity depth, bid--ask spreads, execution quality, and stability. These arise from the interaction between the market design and the behaviour of agents. Finally, some properties are empirical regularities observed in data, such as heavy tails, the volume--volatility relation, and other stylised facts of financial markets (Cont~\cite{cont2001}). From the point of view of AI alignment, this distinction matters: some desirable properties may be built into the mechanism, while others emerge only through the behaviour of agents under the mechanism. Properties such as incentive compatibility and strategy-proofness sit near this boundary: they are behavioural in interpretation, but depend on the logical structure and design of the mechanism.

\section{From market design to emergent failures}

The approach can be illustrated through two canonical market-core mechanisms. The Continuous Limit Order Book represents the familiar sequential view of a market: orders arrive continuously and are processed according to price-time priority. This design is simple and widely used, and it preserves a clear notion of priority and access to liquidity. At the same time, it makes timing and latency structurally important. With the rise of high-frequency trading, speed itself became a strategic advantage: faster participants could exploit tiny timing differences, react first to new information, and extract value through latency arbitrage or the sniping of stale liquidity. These phenomena have been studied in the market-design literature by Budish et al.~\cite{budish2015} and Aquilina et al.~\cite{aquilina2022}.

Frequent Batch Auctions take a different view. Orders arriving within the same time interval are grouped together and treated as time-equivalent, a design proposed as a response to latency races in modern electronic markets~\cite{budish2015}. In this sense, FBA can be interpreted as an attempt to reduce the role of millisecond reaction time and shift competition from speed toward price. This is also connected to the efficient-market ideal that prices should reflect available information, since latency races allow public information to be monetized through tiny speed advantages before prices fully adjust (Fama~\cite{fama1970}). A finer-grained view of FBA exposes the internal steps of batching, activation, matching, and stabilization, making it possible to reason about batching and the computational structure of the mechanism itself.

Similar concerns also arise beyond traditional centralized markets. In decentralized finance, front-running, maximal extractable value, and transaction-ordering manipulation show again that the ordering mechanism is part of the incentive problem~\cite{daian2020,ferreira2023,bartoletti2025}. These examples reinforce the same point: the mechanism that orders, prioritizes, and executes actions is not a neutral implementation detail. It shapes incentives and can create or remove opportunities for strategic exploitation.

This is the sense in which we use the term fundamental market design. The focus is not only on tuning an existing market or measuring its performance after the fact, but on the interaction model that sits at the core of the trading venue. From this perspective, CLOB and FBA are not only market mechanisms; they are different computational answers to the same basic problem: how orders should enter, interact, receive priority, and lead the market toward a stable state.

This also connects to the tradition of agent-based computational economics, where economies are modelled as evolving systems of autonomous interacting agents (Tesfatsion~\cite{tesfatsion2002}). The question becomes sharper when some of these agents are adaptive or AI-based. In such settings, agents may not only pursue fixed strategies; they may learn how the mechanism works and which behaviours it rewards. They may learn to delay, race for speed, withdraw liquidity, or avoid competing too aggressively. These behaviours are not necessarily failures of one agent in isolation. They can arise from the agent--mechanism system.

Several economic failure modes fit this view. Latency races, strategic delay, manipulation of timing or priority, liquidity withdrawal, market unraveling, algorithmic collusion, and performative feedback loops may all emerge from the interaction between agents and rules. These are alignment-relevant because they show that undesirable behaviour may be rational under the wrong mechanism. They also show why monitoring and evaluation should not focus only on agents: we should also ask what incentives the market core creates, what behaviours it makes profitable, and what forms of strategic response it makes possible.

\section{Markets and concurrent computation}

We do not view a market merely as a static order book. An order book is a snapshot of a market at a given moment. A trading venue \textit{is} a computational process. It combines resident liquidity, already present in the market, with incoming order flow. The market core determines how incoming orders interact with the resident state, how priority is applied, which trades are produced, what remains unmatched, and when the market reaches a stable state. This process view is essential for adaptive agents, because such agents do not interact with a static book alone; they interact with a dynamic mechanism over time. The foundational question is therefore: what is the most natural computational model of this interaction from first principles?

There is a further point. The computation taking place on a trading venue is naturally concurrent and parallel: many agents act independently, many orders may be present at the same time, and several potential interactions may be enabled by the same market state. These are principles well understood in computer science, where one distinguishes between independent computations that can proceed in parallel and concurrent processes whose interactions must be coordinated. Yet the dominant market models do not usually start from parallelism and concurrency as fundamental design principles. CLOB resolves interaction through a sequential order-by-order discipline, while FBA introduces batching but is often treated as a clearing rule rather than as a fully explicit concurrent computation. From our perspective, one should ask what the basic parallel and concurrent primitives of a trading venue ought to be: which interactions can proceed independently, which must be ordered by priority, and which synchronization points are needed to preserve stability and transparent execution.

This is also why simulation alone is not enough. Simulation is valuable, especially for exploring large-scale behaviour, but it samples executions of a mechanism. It does not by itself characterize the space of behaviours the mechanism enables. In open, heterogeneous, concurrent markets, the number of possible executions can grow very rapidly because of process interleavings, nondeterminism, agent choices, and changing market state. Formal models can complement simulation by exposing the mechanism that generates behaviour, by making assumptions explicit, and by supporting reasoning about classes of executions rather than only observed runs.

\section{Toward aligned market mechanisms}

More broadly, this perspective matters for AI systems that act in structured social and economic settings. A central lesson from the intersection of artificial intelligence and economic reasoning is that intelligent behaviour cannot be studied apart from incentives, markets, and institutions~\cite{parkes2015}. In financial systems, agents act through mechanisms that define what can be done, which actions interact, and which outcomes are possible. This becomes especially relevant when markets are populated by many algorithmic or AI agents, because the mechanism is the infrastructure through which their interactions are~organised.

The goal is to design transparent market cores that support many-agent interaction while preserving priority, stability, and transparent execution. Parallelism and concurrency are important in this perspective, but only under discipline: a good market core should allow independent interactions to proceed when they can, while making explicit where priority, synchronization, and stabilization are required. Otherwise, parallelism itself can become a source of strategic advantage rather than an improvement in market quality. Transparent formal market cores could also support monitoring, robustness analysis, and incentive-oriented evaluation. In this sense, the task is not only to align the agent, but also to align the market through which agents~interact.

Our position is centered on a simple point: before studying how AI agents behave in markets, we should make the mechanisms through which they interact explicit, transparent, and available for formal~reasoning.

A core designed from better first principles should preserve the strengths of existing models while addressing their weaknesses. CLOB supports smooth price discovery, but its sequential structure makes speed too central. FBA reduces the speed race, but arguably trades off some continuity of price discovery. These limitations are not only implementation issues; they reflect fundamental design choices in the market core. Future AI-driven markets require not only aligned agents, but also aligned mechanisms.

\end{document}